\documentclass[
  submission,copyright,creativecommons,
  colorlinks,allcolors=Blue,
  status=final]
  {eptcs}

\usepackage{graphicx}
\usepackage[svgnames,table]{xcolor}
  \definecolor{lightgray}{gray}{0.9}
\usepackage{url}
\usepackage[english]{babel}
\usepackage{amsmath,amssymb}
\usepackage{paralist}
\usepackage[font={small}]{caption,subfig} 
\usepackage{tikz}
  \usetikzlibrary{arrows,positioning,automata,
    shapes,decorations.pathreplacing,calc}
\usepackage{mdframed}

\newcommand{\wrts}{w.r.t.\ }
\newcommand{\cf}{cf.\ } 		
\newcommand{\ies}{i.e.,\ }

\newcommand{\egs}{e.g.\ } 		
\newcommand{\Eg}{For example,\ }

\usepackage{fixme}
\usepackage{ctable} 
\usepackage{cleveref} 
  \Crefname{section}{Sec.}{Secs.}
  \Crefname{table}{Tab.}{Tabs.}
  \Crefname{figure}{Fig.}{Figs.}
  \Crefname{equation}{Eq.}{Eqs.}
  \Crefname{claim}{Claim}{Claims}

\newtheorem{claim}{Claim}
\newmdtheoremenv[%
hidealllines=true,
leftmargin=20,%
rightmargin=20,
backgroundcolor=gray!30,%
ntheorem]{example}{Running Example Pt.}

\newcommand{\sem}[1]{[\hspace{-.15em}[\mathit{#1}]\hspace{-.15em}]}
\newcommand{\act}[1]{\mathsf{#1}}
\newcommand{\proc}[1]{\mathit{#1}}
\newcommand{\decision}[1]{\mathit{#1}}
\newcommand{\paract}[1]{\mathit{#1}}
\newcommand{\seqcomp}{;}
\newcommand{\altcomp}{\mid}
\newcommand{\parcomp}{\parallel}
\newcommand{\repcomp}[1]{#1^*}
\newcommand{\ControlLoop}{\mathcal{L}}
\newcommand{\AtomSit}[1]{$\act{#1}$}

\newcommand{\stmis}[1][]{\mathit{mis}}

\newcommand{\ac}[2][]{\mathit{#2}_{#1}}
\newcommand{\acendg}[2][]{\ac[#1]{e}^{\mathit{#2}}}
\newcommand{\accomp}[2][]{\ac[#1]{m}^{\mathit{#2}}}

\newcommand{\CtrLoop}{\mathcal{L}}

\newcommand{\RS}{\mathfrak{R}}

\newcommand{\cfct}{\mathit{cf}}
\newcommand{\Yap}{\textsc{Yap}}

\providecommand{\event}{FVAV 2017} 
\usepackage{breakurl}             
\usepackage{underscore}

\title{Run-Time Risk Mitigation in Automated Vehicles:\\
  A Model for Studying Preparatory Steps}
\author{Mario Gleirscher
\institute{}
\institute{Faculty of Informatics\\
Technical University of Munich\\
Munich, Germany}
\email{mario.gleirscher@tum.de}
}
\def\titlerunning{Run-Time Risk Mitigation}
\def\authorrunning{M. Gleirscher}
\begin{document}
\maketitle

\begin{abstract}
  We assume that autonomous or highly automated driving (AD) will be
  accompanied by tough \emph{assurance obligations} exceeding the
  requirements of even recent revisions of ISO 26262 or SOTIF.  Hence,
  automotive control and safety engineers have to
  \begin{inparaenum}[(i)]
  \item comprehensively analyze the driving process and its control
    loop,
  \item identify relevant hazards stemming from this loop,
  \item establish feasible automated measures for the effective
    mitigation of these hazards or the alleviation of their
    consequences.
  \end{inparaenum}

  By studying an example, this article investigates some achievements
  in the modeling for the steps (i), (ii), and (iii), amenable to
  formal verification of desired properties derived from potential
  assurance obligations such as the \emph{global existence of an
    effective mitigation strategy}.  In addition, the proposed
  approach is meant for step-wise refinement towards the automated
  synthesis of 
  AD safety controllers implementing such properties.
\end{abstract}

\section{Introduction}
\label{sec:intro}

For many people, driving a car is a difficult task even after guided
training and many years of driving practice: This statement gets more
tangible when driving in dense urban traffic, complex road settings,
road construction zones, unknown areas, with hard-to-predict traffic
co-participants (\ies cars, trucks, cyclists, pedestrians), bothersome
congestion, or driving a defective car.  Consequently, hazards such as
drivers misjudging situations and making poor decisions have had a
long tradition.  Hence, road vehicles have been equipped with many
sorts of safety mechanisms, most recently, with functions for ``safety
decision support''~\cite{Eastwood2013}, driving assistance, and
monitor-actuator designs, all aiming at reducing operational risk or
making a driver's role less critical.  In AD, more highly automated
mechanisms will have to contribute to risk mitigation at run-time and,
thus, constitute even stronger assurance
obligations~\cite{Eastwood2013}.

In \Cref{sec:background}, we introduce basic terms for \emph{risk
  analysis and (run-time) mitigation~(RAM)} for automated
vehicles~(AV) as discussed in this work.  \Cref{sec:motivation}
elaborates some of these assurance obligations.

\subsection{Background and Terminology}
\label{sec:background}
\label{sec:terminology}

According to control theory, a \emph{control loop} $\mathcal{L}$
comprises a (physical) \emph{process} $P$ to be controlled and a
\emph{controller} $C$, \ies a system in charge of controlling this
process according to some laws defined by an application and an
\emph{operator}~\cite{Lunze2010}.  The engineering of safety-critical
control loops typically involves reducing \emph{hazards} by making
controllers safe in their intended function (SOTIF), resilient (\ies
tolerate disturbances), dependable (\ies tolerate faults), and secure
(\ies tolerate misuse).

In automated driving, the process under consideration is the
\emph{driving process} which we decompose into a set of \emph{driving
  situations} $\mathcal{S}$, see \Cref{sec:sitmodel}.  Taxonomies of
such situations have been published in, \egs~\cite{TSC2017}.
Based on recommendations of SAE and U.S.~DoT
\cite{ORADC2016,USDOT2016}, we distinguish the following \emph{modes}
or \emph{levels}~(L) of automation for AD: AD assistance (ADAS, L1-2),
highly automated (HAD, L3-4), and fully automated or autonomous (FAD,
L5).  Yet, in level 4, a human driver is supposed to stay ``weakly''
in-the-loop for an occasional machine-to-human hand over.

By the term \emph{causal factor} (CF), we denote any concept ranging
from a \emph{root cause} to a \emph{hazard}, to a
\emph{near-mishap}.\footnote{\Eg a road accident for which airbags
  successfully alleviated harm.}  Causal factors can form causal
chains describing effect propagation through $\ControlLoop$,
particularly, a near-mishap can lead to another root
cause.\footnote{This helps fixing the ``Swiss cheese problem'' raised
  by \textsc{Leveson} \cite{Leveson2012}.}  By \emph{causal factor
  model} (also: hazard or risk model), we refer to the collection of
causal factor descriptions used for a RAM application.

A \emph{rare undesired event} (RUE, also: hazardous event or
situation) denotes any \emph{state of $\mathcal{L}$}
\begin{inparaenum}[(i)]
\item with an occurrence of one or more causal factors and
\item arguably increasing the risk of mishaps.
\end{inparaenum}
Moreover, by \emph{(operational) risk}~\cite{LundSolhaugStoelen2011},
we refer to the likelihood---quantitatively, the probability or
frequency---of reaching a specific set of RUEs from any current state.
Then, a \emph{safe state} is any state with acceptably low
risk---quantitatively, risk below a \emph{budget} $b$.  Although we
use further information for the modeling in RAM, any safe state can be
transcribed into a corresponding behavioral invariant or a safety
property of $\mathcal{L}$ in the sense of
\cite{DBLP:journals/tse/Lamport77,Manna1995}.
  
\begin{figure}
  \centering
  \subfloat[]{
    \begin{tikzpicture}
      \input{figures/riskstructure-phases}
    \end{tikzpicture}
    \label{fig:hazardphasemodel-generic}
  }
  \qquad
  \subfloat[]{
    \includegraphics[height=4cm]{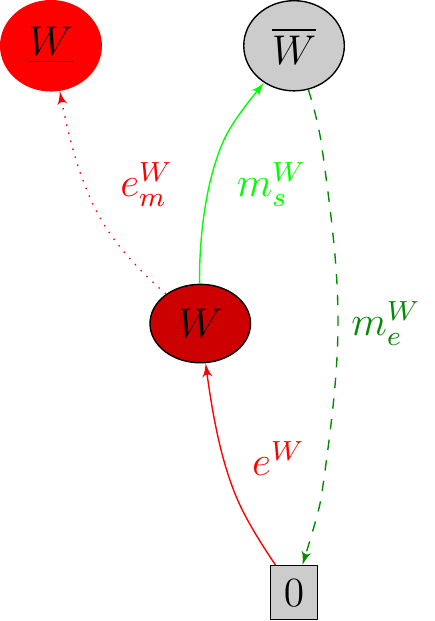}
    \label{fig:hazardphasemodel-badweather}
  }
  \caption{Generic phase model for causal factors (a) exemplified for
    the causal factor $W$ for \texttt{badWeather} (b).}
  \label{fig:hazardphasemodel}
\end{figure}

Elaborating on the notion of \emph{behavioral safety} in
\cite{Gleirscher2014a}, we consider two types of actions in
$\mathcal{L}$: \emph{endangerments} are actions whose performance
leads to RUEs, and \emph{mitigations} are actions whose performance
represents countermeasures for specific sets of RUEs.
For any causal factor $\cfct$, we consider two types of endangerments:
$e^{\cfct}$ denotes activation resulting in a change to phase
$\cfct$ (\ies \emph{active}), and $e_m^{\cfct}$ describes any
\emph{mishap} $\underline{\cfct}$ potentially following
${\cfct}$.  Furthermore, we consider three types of mitigations:
$m_s^{\cfct}$ initiates mitigation by a change to phase
$\overline{\cfct}$ (\ies \emph{mitigated}), $m_e^{\cfct}$ completes
mitigation by a change to phase $0^{\cfct}$ (\ies \emph{inactive}),
and $m_c^{\cfct}$ directly deactivates $\cfct$ and completely restores
its consequences, again by a change to $0^{\cfct}$.
Shown in \Cref{fig:hazardphasemodel}, these notions lead to what we
call \emph{phase model}\footnote{Gray transitions indicate that we
  might employ more expressive variants of this phase model as, \egs
  discussed in \cite{DBLP:conf/hase/KumarS17} for a different context.
  Operationally, the phase model would perform non-observable (also:
  silent) transitions if none of the described actions is enabled.
  Similar abstractions are used for testing, \egs in
  \cite{DBLP:conf/fortest/Tretmans08}.
} \cite{Gleirscher2017-NFM}. Phase models can be composed.
Then, by $0$ we refer to the ``safest'' state in a model composed of
phase models for several causal factors, see
\Cref{sec:riskfactorization}.

For the description of these concepts, we employ \emph{labeled
  transition system (LTS)}
modeling. 
An LTS is a tuple $(\Sigma,\mathcal{A},\Delta)$ with a set of
\emph{states} $\Sigma$, a set of \emph{action} labels $\mathcal{A}$,
and a transition relation
$\Delta\subseteq \Sigma\times\mathcal{A}\times\Sigma$.  For a
\emph{transition} $(\sigma,\alpha,\sigma')\in\Delta$, we also say that
the state $\sigma'$ is the \emph{event} observed from the performance
of the action $\alpha$.
We will work with the (usually dense) \emph{loop state space}
$\Sigma_{\mathcal{L}}$, the finite \emph{situation state space}
$\Sigma_{\mathcal{S}}$, and the finite \emph{risk state space}
$\Sigma$.  The symbols used to model CFs in $\Sigma$
(\Cref{fig:hazardphasemodel}) can refer to predicates over
$\Sigma_{\mathcal{L}}$. 
Further details on this terminology will follow in
\Cref{sec:sitmodel,sec:riskfactorization}.

\subsection{Motivation and Challenges}
\label{sec:motivation}

Let us take the viewpoint of a safety engineer dealing with the
assurance of AVs not only compliant with available standards (\egs ISO
26262) but also trying to achieve the level of safety desired by any
traffic participant.  What does that mean?  Well, our engineer might
try to argue towards the following claim from manual driving:
\begin{claim}
  \label{thm:toplevelsafetygoal}
  The rational driver always tries to reach and maintain a safe state
  \wrts all RUEs recognizable and reacted upon by the driver in any
  driving situation.
\end{claim}
The reader might not agree with this claim or require further claims
to be established in this context.  However, for sake of simplicity,
the following discussion is restricted to this claim.
Next, we list three \textbf{tasks} of our safety engineer responsible for
RAM for AD:
\begin{description}
\item[Task S:] investigate the domain of \emph{situations} in the driving
  process,
\item[Task E:] investigate the domain of \emph{endangerments}, identify
  causal factors and potential mishaps,
\item[Task M:] investigate the domain of \emph{mitigations}, prepare
  effective countermeasures.
\end{description}
These tasks are error prone, largely subject to expert judgment and
require rigorous method because
\begin{inparaitem}
\item driving is a complex task (see beginning of \Cref{sec:intro})
\item we aim at FAD,\footnote{%
    See, \egs \url{http://www.cnbc.com/2017/06/07/robot-cars-cant-count-on-us-in-an-emergency.html}.
  }
\item we might need to automate mitigations as well,\footnote{%
    See, \egs
    \url{https://www.theguardian.com/technology/2017/apr/04/uber-google-waymo-self-driving-cars}.
  }
\item we have to \emph{argue} that the measures developed (\textbf{M}) can
  mitigate the RUEs identified (\textbf{E}) in the considered
  driving process (\textbf{S}) and residual risk
  stays acceptably low, \egs below an upper bound $b$.
\end{inparaitem}
We investigate these tasks using the notions from
\Cref{sec:terminology}, discuss some challenges as well as a RAM
approach based on this model:

For \textbf{Task S}, we have to come up with a model of driving
situations reflecting the driving process in the road environment (see
\Cref{sec:sitmodel}), 
abstract but comprehensive
enough for RAM purposes.

\begin{example}
  Throughout the paper, we will work with
  an example highlighting aspects of RAM for AD.  
  From our knowledge about the driving domain, we can quickly derive a
  large number of situations such as, \egs leaving a parking lot,
  entering a highway, halting somewhere, driving through road
  construction zones, overtaking in urban traffic, and, more
  generally, AD at level 4, called \AtomSit{drivingAtL4Generic} below.
\end{example}

For the automation of some analyses, we use a newly developed tool
called \Yap\ based on concepts discussed in
\cite{Gleirscher2014a,Gleirscher2017-NFM}.  \Yap\ stands for ``yet
another planner.'' The complete example for this paper, a preliminary
version of \Yap, and a user's manual are available online.\footnote{%
Demonstration artifacts can be downloaded from \href{http://gleirscher.de}{http://gleirscher.de}.}

\emph{Questions from Task S:}
(S1) Which situations form equivalence classes?
(S2) Which situations do we have to discriminate in our model?
(S3) How are the discriminated situations related?

For \textbf{Task E}, given a set of situations from Task S, we have to
understand how a \emph{safe state} in each situation looks like as
well as all the undesirable ways of leaving this state,\footnote{For
  behavioral safety we have to deal with leaving safety invariants in
  many ways.  Our interest, hence, lies on accepting to leave an
  ``inner invariant,'' \ies allowing causal factors to occur, and
  trying not to leave an ``outer invariant,'' \ies reducing the
  likelihood of an accident.  This way, we employ \textsc{Lamport}'s
  \cite{DBLP:journals/tse/Lamport77} notion of safety property in a
  layered manner, similar to the discussion of ``safety envelopes''
  \cite{Koopman2016} or layer-of-protection analysis
  (LOPA)~\cite{Dowell1998}.} \ies endangerments leading to some RUE.
Note that RUEs can model simultaneously occurring causal factors.

\begin{figure}
  \includegraphics[viewport=6000 0 7750 600,clip,width=\textwidth]{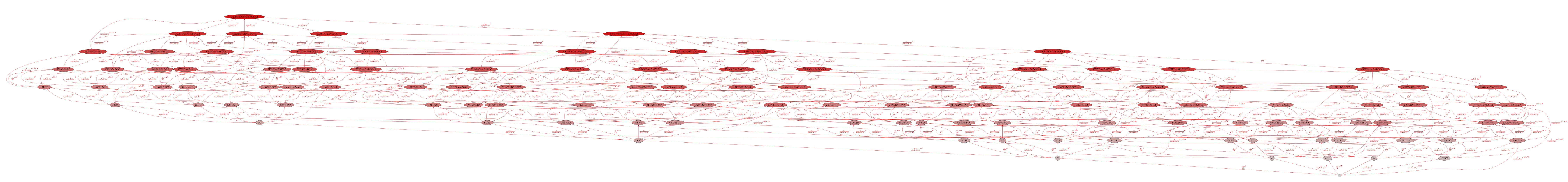}
  \caption{Cutout of the risk state space $\Sigma$ indicating endangerment complexity:
    \emph{nodes} indicate RUEs to be regarded in a situation;
    \emph{edges} indicate potential endangerments leading to
    these RUEs.}
  \label{fig:os-driveAtL4generic-endgonly}
\end{figure}

\begin{example}
Safety engineers of the driving domain would
derive many causal factors such as, \egs sensor fault, inattentive
driver, electric power shortage, low fuel.
For the situation \AtomSit{driveAtL4Generic},
\Cref{fig:os-driveAtL4generic-endgonly} depicts risk complexity in
terms of RUEs hypothetically reachable and predictable from the
state $0$ for which we might be required to provide mitigations in
any AD mode.
Combining 10 causal factors (incl.\ the mentioned ones) using the
phase model from \Cref{sec:background}, with initial state $0$, we
get 7128 states reachable via 51984 transitions.
By only regarding ways of how we directly get into a RUE, we are
left to consider 112 states and 372 transitions.
Further details see \Cref{sec:riskfactorization,sec:runtimemitigation}.
\end{example}

\emph{Questions from Task E:}
(E1) Which causal factors form equivalence classes?
(E2) Which ones do we have to include in our model?
(E3) How can we further classify these factors?
(E4) Which factors have to be focused?
(E5) How are the remaining factors related to each other?

For \textbf{Task M}, given the causal factors from Task E, we have to
select and prepare optimal ways of how each of the RUEs can be
mitigated.

\emph{Questions from Task M:}
(M1) Which mitigation is required for which RUE in which situation?
(M2) How can we represent this in our model?
(M3) Does it help to classify RUEs according to the available
mitigations?
(M4) Can we construct mitigations applicable to classes of RUEs to get
a minimal set of effective mitigations for a given set of situations?

Aligned with the challenges summarized in \cite{Guiochet2017}, the
questions S1-3, E1-5, and M1-4 define our general research setting.

\paragraph{Contributions and Outline.}
\label{sec:contributions}

We show how we get to a model for RAM for AD in a
systematic way, particularly, keeping a link to the detailed models
used by control engineers for the controlled process.  We discuss
\begin{inparaitem}
\item 
  the modeling of the driving process as a controlled process over
  driving situations (Task S, \Cref{sec:sitmodel}),
\item 
  a refinement of the transition system model discussed in
  \cite{Gleirscher2017-NFM},
\item 
  the classification and modeling of multiple causal factors
  comprising RUEs in the AD control loop specific to a set of driving
  situations (\Cref{sec:hazid,sec:riskfactorization}),
\item the state space exploration based on this model together with
  constraints for reducing and shaping this state space (Task E,
  \Cref{sec:statespacereduction-endg,sec:statespacereduction-mit}),
\item 
  a taxonomy of mitigation actions (\Cref{sec:mitigationtaxonomy}),
  and
\item the usage of this model for RAM (Task M,
  \Cref{sec:runtimemitigation}).
\end{inparaitem}
We discuss related work in \Cref{sec:relwork}, further issues and a
research vision in \Cref{sec:further-discussion}, and conclude in
\Cref{sec:conclusions}.

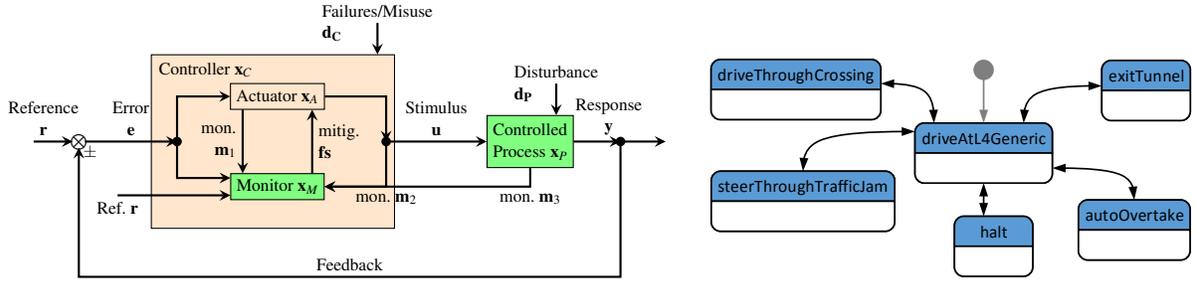
\begin{figure}
  \subfloat[The generic control loop $\mathcal{L}$ including (i) a
  ``monitor'' acting as our \emph{safety controller} and \emph{mitigation
    planner} and (ii) the controlled process, in our case ``driving.'']{
    \begin{tikzpicture}
  [->,>=stealth,
  scale=.6,every node/.style={transform shape},
  box/.style={draw,align=center,minimum height=3em}]


  \node[box,fill=orange!20,minimum width=14em,minimum height=10em]
  (ctr) at (2.3,0) {};
  \node[anchor=north west] at ($(ctr.north west)+(.1em,-.1em)$) {Controller $\mathbf{x}_C$};
  \node[box,fill=green!50] (proc) at (8,0) {Controlled\\Process $\mathbf{x}_P$};

  \node[draw] (actuator) at ($(ctr)+(.1,1)$) {Actuator $\mathbf{x}_{A}$}; 
  \node[draw,fill=green!50] (monitor) at ($(ctr)+(.1,-1)$) {Monitor $\mathbf{x}_{M}$}; 
  
  \node (error) at (-2,0) {}; 
  \node at (error) {$\bigotimes$};
  \node at ($(error.south east)+(.3em,-.3em)$) {$\pm$};
  \coordinate (jct) at ($(proc)+(2,0)$); 
  \fill (jct) circle (3pt);
  \coordinate (jctmon) at ($(ctr.east)+(-.5em,0)$); 
  \fill (jctmon) circle (3pt);
  \coordinate (jctinput) at ($(ctr.west)+(1.5em,0)$); 
  \fill (jctinput) circle (3pt);

  \draw[thick,->] 
  ($(actuator.south)+(-2em,0)$) -- node[left,align=right] {mon.\\$\mathbf{m}_1$} ($(monitor.north)+(-2em,0)$);
  \draw[thick,->] 
  ($(monitor.north)+(2em,0)$) -- node[right,align=left]{mitig.\\
    $\mathbf{fs}$} ($(actuator.south)+(2em,0)$);
  \draw[<-,thick] 
  (error) 
  edge[above,align=center] 
  node[near end] {Reference\\$\mathbf{r}$} 
  ++(-1,0);
  \draw[thick,->] 
  (jct) -- ++(0,-3) -- node[above] {Feedback}
  ($(error)+(0,-3)$) -- (error);
  \draw[thick,->] (jctinput) |- (actuator);
  \draw[thick,->] (jctinput) |- ($(monitor.west)+(0,.5em)$);
  \draw[thick,->] (jctmon) |- node[below] {mon.\ $\mathbf{m}_2$} (monitor);
  \draw[thick,->] (proc) |- node[below] {mon.\ $\mathbf{m}_3$} (monitor);
  \draw[<-,thick] 
  ($(monitor.west)+(0,-.5em)$) 
  -- node[below,at end] {Ref.\ $\mathbf{r}$} 
  ++(-2.5,0);
  \draw[->,thick] 
  (actuator) -| 
  (jctmon) edge node[above,align=center] {Stimulus\\$\mathbf{u}$} (proc)
  (proc) edge node[near end,above,align=center] {Response\\$\mathbf{y}$} (jct)
  (error) edge node[above,align=center] {Error\\$\mathbf{e}$} (jctinput);
  \draw[->,thick] (jct) -- ++(.5,0) -- ++(.5,0);

  \path[draw,arrows={<-},thick] 
  ($(ctr.north east)+(-1em,0)$) edge node[at end,align=left] {Failures/Misuse\\$\mathbf{d_{C}}$} ++(0,.7)
  ($(proc.north east)+(-1em,0)$) -- node[at end,align=left]
  {Disturbance\\$\mathbf{d_{P}}$} ++(0,.7);
\end{tikzpicture}
    \label{fig:controlledprocess-blackbox}
  }
  \hfill
  \subfloat[Driving situations related to the situation
  \AtomSit{drivingAtL4Generic}.]{
    \includegraphics[width=.4\textwidth]{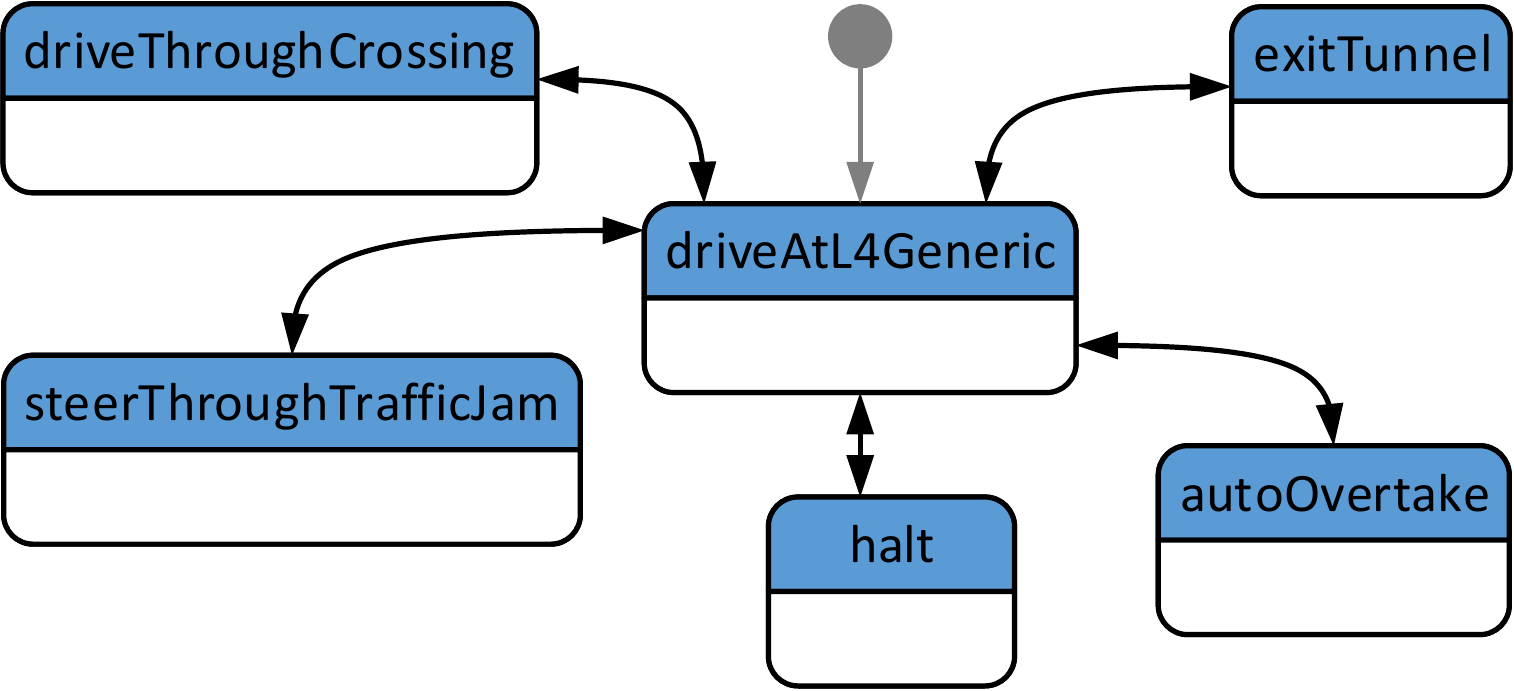}
    \label{fig:osdecisionprocess-cutout}
  }
  \caption{Generic structure of the
    control loop~(a) and cutout
    of the abstraction of the controlled
    process~(b).}
\end{figure}

\section{Modeling Driving Situations, Processes, and Scenarios}
\label{sec:sitmodel}

Based on \Cref{sec:terminology}, we provide an abstraction of the
driving process in $\mathcal{L}$, and a notion of driving scenario for
the symbolic execution of this process as shown in
\Cref{sec:situationchanges}.

\subsection{Preliminaries on Driving Processes}
\label{sec:formal-preliminaries}

\paragraph{State of the Control Loop.}
Given the process output $\mathbf{y}$, reference $\mathbf{r}$, and
control input $\mathbf{u}$, the \emph{process} $P$ (open loop) can be
described by $\mathbf{y}_{new} = P(\mathbf{y}, \mathbf{u})$.  We close
this loop using $\mathbf{u} = C(\mathbf{e})$ for the controller $C$,
$\mathbf{e} = \mathbf{r} - \mathbf{y}_{new}$ for the error $e$.
At this level, we have the state
$(\mathbf{r},\mathbf{u},\mathbf{y}) \in \Sigma_{\CtrLoop} \triangleq
\mathbb{R}^{dim(r)+dim(u)+dim(y)}$ of the control loop
$\CtrLoop$~\cite{Lunze2010}.  However, as shown in
\Cref{fig:controlledprocess-blackbox}, based on the
controller-internal state $\mathbf{x}_C$, controller-related causal
factors $\mathbf{d}_C$, process-internal state $\mathbf{x}_P$, and
process-related causal factors $\mathbf{d}_P$, we get
$\mathbf{u} = C(\mathbf{x}_C, \mathbf{e}, \mathbf{d}_C)$ and state
$ (\mathbf{r}, \mathbf{x}_C, \mathbf{d}_C, \mathbf{u}, \mathbf{x}_P,
\mathbf{d}_P)$ which we can refine \cite{Broy2001} to get
$\mathbf{u} = A(\mathbf{x}_A, \mathbf{e}, \mathbf{fs}, \mathbf{d}_C)$
for the \emph{actuator} $A$ with
$\mathbf{fs} = M(\mathbf{x}_M, \mathbf{m}_{1,2,3}, \mathbf{r},
\mathbf{e})$ for the \emph{monitor} $M$.  For the process, we then
consider
$\mathbf{y}_{new} = P(\mathbf{x}_P, \mathbf{u}, \mathbf{d}_P)$ with
the \emph{loop state}
$\sigma_{\mathcal{L}}=(\mathbf{r}, \mathbf{x}'_C, \mathbf{x}_A,
\mathbf{x}_M, \mathbf{d}_C, \mathbf{u}, \mathbf{x}_P, \mathbf{d}_P)$
where $\mathbf{x}_P$ includes $\textbf{y}$ and
$\textbf{x}_C=(\textbf{x}'_C,\textbf{x}_A,\textbf{x}_M)$.

\paragraph{Driving Processes.}
\label{sec:drivingprocess}
Inspired by process algebra (see, \egs\cite{Milner1995,Hoare1985}), for a
set of \emph{driving situations} $\mathcal{S}$, we represent
\emph{driving processes} $S$ or $T$ by expressions of the form
\begin{equation}
  \label{eq:proccalc}
  S,T ::= s \mid S{\parallel}T \mid S|T \mid S;T \mid S^*\qquad\text{(ordered
    by operator precedence)}
\end{equation}
%
where $s\in\mathcal{S}$, $S;T$ denotes the \emph{sequential
  composition} of $S$ and $T$, $S|T$ \emph{non-deterministic choice}
between $S$ and $T$, $S{\parallel}T$ their \emph{parallel
  composition}, and, for convenience, $S^*$ the (possibly empty)
\emph{repetition} of $S$.  Given the set $\mathcal{S}^*$ of
tuple\footnote{We use tuples to represent parallel composition.}
sequences over $\mathcal{S}$, a process $S$ defines a set of
\emph{driving scenarios} (also: situation traces or process runs)
denoted by $\sem{S}$ and defined by
\begin{align*}
  \sem{S} &= \{S\} \quad\Leftrightarrow S\in\mathcal{S} &
  \sem{S\seqcomp\proc{T}} &= \{ st\in\mathcal{S}^* \mid
                            s\in\sem{S}\land t\in\sem{T}\} && \\
  \sem{S\altcomp\proc{T}} &= \sem{S}\cup\sem{T} &
  \sem{S\parcomp\proc{T}} &= \{ (s,t)\in\sem{S}\times\sem{T}\mid
                        (s,t)\models \Phi \} &
  \sem{\repcomp{S}} &= \emptyset\cup\sem{S\seqcomp\repcomp{S}}
  \label{eq:procsem}
\end{align*}
where $\Phi$ is a formula used to constrain $\sem{S\parcomp T}$ to
admissible sequences of situation tuples in $\mathcal{S}^*$. A
discussion of algebraic process properties and instances of $\Phi$
would go beyond the scope of this paper.
  
We visualize a \emph{driving process} $S$ by a graph: Nodes represent
situations and edges the successor relation for envisaged transitions
between such situations.  Bidirectional edges abbreviate two
correspondingly directed edges.  The anonymous initial situation is
indicated by a gray dot.  Such a graph can visualize the execution of
\emph{driving scenarios} compliant with the successor relation.

\begin{example}
\Cref{fig:osdecisionprocess-cutout} shows driving situations related
to situation \AtomSit{drivingAtL4Generic}.  Note that the model in
\Cref{fig:osdecisionprocess-cutout} simplifies reality, \egs
\AtomSit{halt} can follow \AtomSit{autoOvertake} only indirectly after
visiting \AtomSit{driveAtL4\-Gene\-ric} for an instant.  We do not
consider this as a restriction of expressiveness.
\end{example}

\subsection{Identification of Hazards for a Driving Situation}
\label{sec:hazid}

Knowing a fragment of the driving process, we can perform hazard
identification and selection for the situations we identified.  We
assume that early-stage and process-level forward and backward
analysis~\cite{Ericson2015} can be applied, \egs domain-specific
checklists, hazard identification (HazId) or analysis (HazAn), hazard
operability study (HazOp), failure-mode and effects (FMEA), event-tree
(ETA), fault-tree (FTA), cause-consequence (CCA)~\cite{Nielsen1971},
or 
system-theoretic process (STPA)~\cite{Leveson2012} analysis.

\begin{example}
For the situation \AtomSit{driveAtL4Generic}, we conducted a
light-weight, HazOp-like forward and backward analysis resulting in 10
causal factors, \egs
\begin{itemize}
\item \texttt{noAutoPilot} ($nAP$), denoting the class of
  failures 
  of an ``autopilot'' \emph{function}, and
\item \texttt{nearCollision} ($nC$), denoting the event of a
  near-collision.
\end{itemize}
Based on $\sigma_{\mathcal{L}}$ (\Cref{sec:formal-preliminaries}),
\texttt{nearCollision} can be a predicate relating \emph{longitudinal
  and angular acceleration} and \emph{direction} of the AV and
\emph{distance} between the AV and other objects nearby.
\end{example}

\subsection{Grouping and Composing Driving Situations}
\label{sec:composedsituations}

Carrying on with process analysis can lead to a large $\mathcal{S}$
representing the driving process for which we have to perform RAM.
Hence, we group situations using regions.  We call a process $S$ a
\emph{region} (also: aspect) if it superimposes (by $\parallel$)
properties of a causal factor model on a subset of $\mathcal{S}$.
Grouping criteria can be
\begin{inparaenum}[(i)]
\item specific modes of vehicle operation, \egs the level of
  automation at which the vehicle is operated (see
  \Cref{sec:terminology}),
\item situations across whom similar AD functions are used, \egs
  \AtomSit{requestTakeOverByDr},
\item equivalence classes over process parameters, \egs
  $\mathbf{x}_P.\mathtt{speed}$.
\end{inparaenum}

\begin{figure}
  \centering
  \subfloat[Despite ``$\parcomp$'' precedes
  ``$\altcomp$'' and ``$\altcomp$'' precedes ``$\seqcomp$'' we use
  parentheses for clarity.]{
    $\begin{aligned}
      \proc{P0} &\equiv
                  \act{start}\seqcomp(\act{basic}\parcomp\act{supplyPower}\parcomp\decision{D1})\\
      \decision{D1} &\equiv
                      (\paract{PP1}\parcomp\act{autoLeaveParkingLot})\seqcomp\proc{P1}\altcomp
                      (\paract{PP1}\parcomp\act{leaveParkingLot})\seqcomp\proc{P2}\\
      \proc{P1} &\equiv
                  (\paract{PP2}\parcomp\act{driveAtL4Generic})\seqcomp\decision{D11}\\
      \proc{P2} &\equiv
                  (\paract{PP3}\parcomp\act{driveAtL1Generic})\seqcomp\decision{D12}\\
      \decision{D11} &\equiv
                       (\paract{PP2}\parcomp(\act{driveThroughCrossing}\altcomp
                       \act{exitTunnel}\altcomp
                       \act{autoOvertake}))\seqcomp(\proc{P1}\altcomp\proc{P2})\altcomp\\
                &\qquad (\paract{PP2}\parcomp\act{halt})\seqcomp
                  (\proc{P1}\altcomp\proc{P2}\altcomp
                  (\paract{PP1}\parcomp\act{parkWithRemote}\seqcomp
                  \decision{D1}))\altcomp\\
                &\qquad (\paract{PP1}\parcomp\act{steerThroughTrafficJam})\seqcomp\proc{P1}\\
      \decision{D12} &\equiv \proc{P1}\altcomp
                       (\paract{PP3}\parcomp\act{manuallyOvertake})\seqcomp\proc{P2}\altcomp
                       (\paract{PP1}\parcomp\act{manuallyPark})\seqcomp\decision{D1}\\
      \paract{PP1} &\equiv \act{driveAtLowSpeed}\\
      \paract{PP2} &\equiv \act{drive}\parcomp
                     \act{driveAtL4}\parcomp\act{requestTakeoverByDr}\\
      \paract{PP3} &\equiv \act{drive}\parcomp
                     \act{driveAtL1}\parcomp\act{operateVehicle}
    \end{aligned}$
    \label{tab:drivingprocess-prodrules}
  }
  \vspace{1em}
  \subfloat[In blue, the \emph{driving
    situations} comprising the controlled process; in gray,
  \emph{regions} grouping properties of the \emph{causal factor
    model} built for this example.]{
    \includegraphics[width=.7\textwidth]{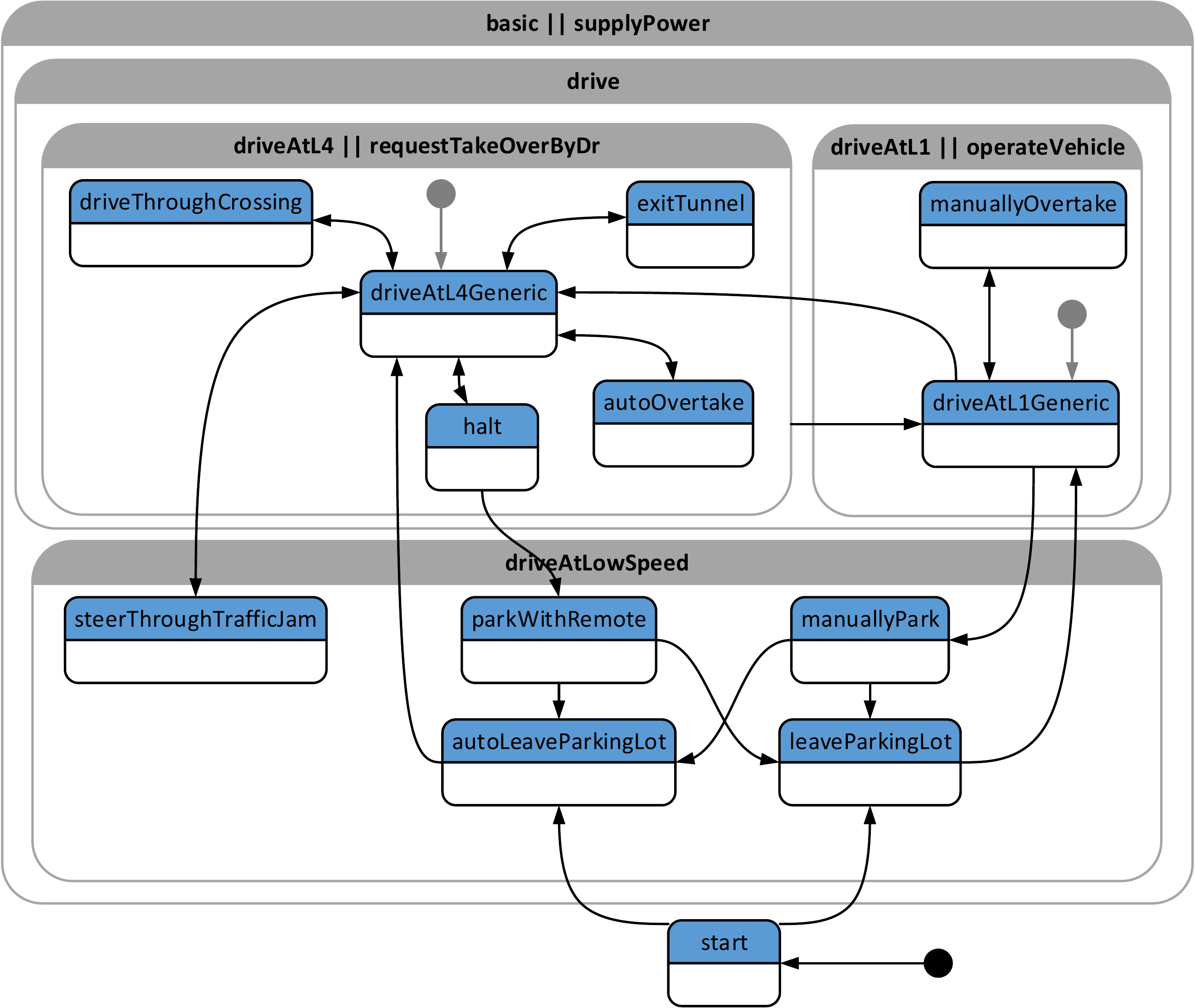}
    \label{fig:osdecisionprocess-overall}
  }
  \caption{A \emph{driving process} constructed according to
  \Cref{eq:proccalc}: recursive expressions~(a) and their visualization~(b).}
\end{figure}

\begin{example}
Following \Cref{eq:proccalc}, \Cref{tab:drivingprocess-prodrules}
models a driving process.  Visualized in
\Cref{fig:osdecisionprocess-overall}, regions are shown in gray.  For
the region containing \AtomSit{supplyPower}, we applied forward
reasoning (\egs FMEA) to power-related vehicle components of the loop.
We identified 3 sub-system failure modes as causal factors:
\begin{itemize}
\item \texttt{lowOrNo-Fuel} ($F$), denoting low fuel level or no more fuel,
\item \texttt{lowOrNo-Energy} ($E$), denoting reduced 
  or outage of primary electric power supply, and
\item \texttt{lowOrNo-Battery} ($B$), denoting reduced or outage of
  secondary electric power supply.
\end{itemize}
The results from FMEA provide information about the phases
$\underline{F},\underline{E}$, and $\underline{B}$ of these phase
models.
\end{example}

\section{What Can Happen in a Specific Driving Situation?}
\label{sec:riskfactorization}

In this section, we will investigate a structure for identifying and
characterizing run-time risk.

\subsection{Risk Structures}
\label{sec:riskstructures}

\begin{figure}[t]
  \subfloat[]{
    \includegraphics[width=.5\textwidth]{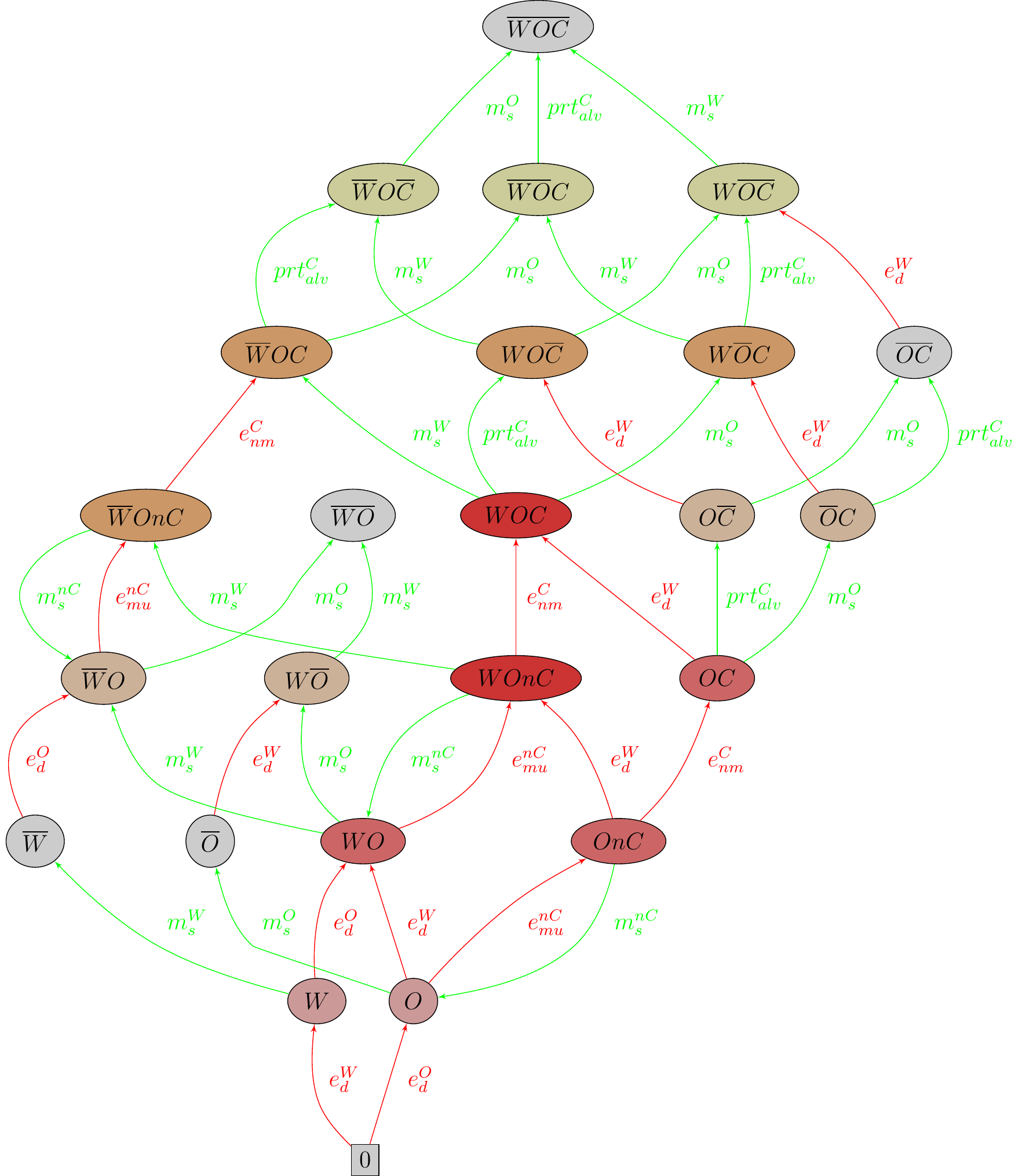}
    \label{fig:os-drive}}
  \subfloat[]{
    \includegraphics[height=.3\textheight]{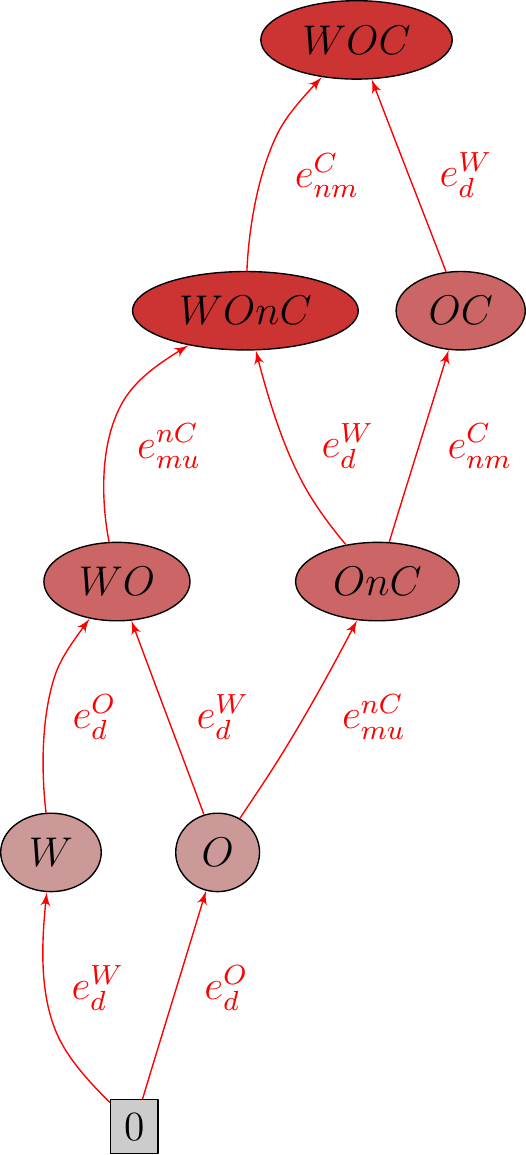}
    \label{fig:os-drive-endgonly}}
  \subfloat[]{
    \includegraphics[height=4cm]{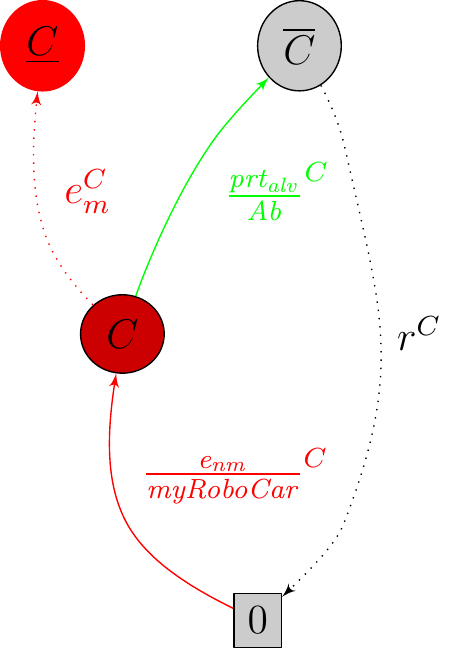}
    \label{fig:os-drive-collision}}
  \subfloat[]{
    \includegraphics[height=4cm]{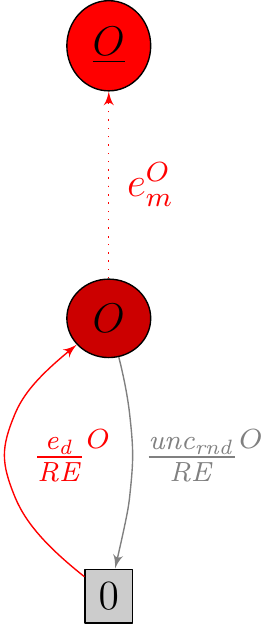}
    \label{fig:os-drive-badweather}}
  \caption{The situation \AtomSit{drive} as an aspect:
    (a) The whole \AtomSit{drive} aspect,
    (b) endangerment paths in \AtomSit{drive} starting from $0$,
    (c) phase model for \texttt{collision} ($C$) and
    (d) for \texttt{obstacleInTrajectory} ($O$).}
  \label{fig:os-graphs}
\end{figure}

For any situation $s\in\mathcal{S}$, we use the LTS model
(\Cref{sec:terminology}) to create a \emph{risk structure}
$\mathfrak{R}_s=(\Sigma_s,\mathcal{A}_s,\Delta_s)$~\cite{Gleirscher2017b,Gleirscher2017-NFM}
from parallel composition of \emph{phase models} for the set of
causal factors factorizing risk in $s$.
The graphs in
\Cref{fig:hazardphasemodel-badweather,fig:os-driveAtL4generic-endgonly,fig:os-graphs}
show applications of the phase model:
\begin{itemize}
\item nodes depict \emph{risk states} $\subseteq\Sigma_s$ composed of
  phases, $0$ denoting the ``safest'' state,
\item red edges are transitions $\subseteq\Delta_s$ to states of
  higher risk (\emph{endangerments} $\subseteq\mathcal{A}_s$),
  particularly, from $0$,
\item green edges are transitions $\subseteq\Delta_s$ to states of
  lower risk (\emph{mitigations} $\subseteq\mathcal{A}_s$),
  preferably, towards $0$.
\end{itemize}
In $\RS_s$, RUEs are states in $\Sigma_s$ related with mishaps of
high severity reached by corresponding combinations of endangerments.
Although we can apply both, backward and forward analysis, to construct
$\RS_s$ using the phase model, this work goes along the lines
of \cite{Gleirscher2017-NFM} and carries on with forward construction.
For RAM, we consider this to be natural: We want to predict
the \emph{step-wise approach of states of higher risk reachable from
  any current state} (usually $0$) and identify early mitigation
steps (\ies points of interception) rather than starting with mishap
analysis (see, \egs \cite{Svedung2002}).

\begin{example}
The situation \AtomSit{drive} represents a common situation composed
in parallel with the situations \AtomSit{basic},
\AtomSit{supplyPower}, and the process
$(\act{driveAtL4}\parcomp\act{requestTakeoverByDr}\altcomp
\act{driveAtL1}\parcomp\act{operateVehicle})^*$ including its
subordinate driving situations.
We can understand \AtomSit{drive} as an \emph{aspect} of these
situations and, consequently, of the \emph{scenarios} composed with
\AtomSit{drive}.  With \AtomSit{drive}, we associate the following
CFs: \texttt{badWeather} $(W)$, \texttt{obstacleInTrajectory} $(O)$,
\texttt{nearCollision} $(nC)$, and \texttt{collision} $(C)$.

\Cref{fig:os-drive} depicts a \emph{constrained composition} of the
phase models for $W,O,nC$, and $C$.  In the following, particularly,
in \Cref{sec:statespacereduction-endg}, we will discuss constraints
that led to $\RS_{\act{drive}}$.
\end{example}

\subsection{Selecting Hazards According to Driving Situations}
\label{sec:hazardselection}

As we have already seen, situations help focus on a specific part of
the controlled process and, consequently, our hazard analysis.
Particularly, situations can induce specific relationships between
CFs.  Moreover, relationships between a fixed pair of CFs can change
from one situation to another.  However, the final decision on the set
of relevant CFs for a specific situation depends on many criteria
addressed by the techniques mentioned in \Cref{sec:hazid}.

\begin{example}
A CF stemming from undesired behavior of human operators only has
influence on a driving situation where an operator is actually in the
loop.  In our model, the causal factor \texttt{dumbDriving} ($D$)
\textbf{causes} $nC$ in any situation composed with the aspect
\AtomSit{operate\-Vehicle}, however, in any situation composed with
the aspect \AtomSit{driveAtL4}, $D$ is not part of the \textbf{causes}
relationship with $nC$.
\end{example}

\subsection{Which Events Increase Risk? What is an
  Endangerment Comprised of?}

Based on the model from
\Cref{sec:terminology,sec:formal-preliminaries,sec:riskfactorization},
risk is assumed to be only increased by \emph{endangerments}.
Endangerments are those actions performed in $\mathcal{L}$ that result
in \emph{activations} of causal factors, \ies exactly those
risk-increasing events we want our \emph{safety controller}
(\Cref{fig:controlledprocess-blackbox}) to observe.

\begin{example}
\Cref{fig:os-drive-endgonly} shows combinations of endangerments in
the situation \AtomSit{drive} for which the AV has to be prepared,
\egs when in $0$.  For example, the action $e^O_d$ performs the
activation of the causal factor $O$.  Index $d$ denotes that this
action belongs to the class of disturbances.
\end{example}

\paragraph{Classifying Endangerments and Causal Factors.}
\label{sec:hazardtaxonomy}

We classify a CF by the type of action that activates it,
associated with the part of $\mathcal{L}$ which performs
this action.  For this, we distinguish between
\begin{itemize}
\item \emph{failures} $(f)$, stemming from technical parts of the
  controller,
\item \emph{disturbances} $(d)$, stemming from some parts of the
  controlled process, 
\item \emph{misuse} $(mu)$, stemming from anyone who has access to the
  controls (\egs human operator), and 
\item \emph{near-mishap} ($nm$), a special class used to model
  endangerments that lead to what we understand as ``a mishap for which
  we are able to prepare effective countermeasures.''
\end{itemize}

\begin{example}
In \Cref{fig:os-drive-endgonly}, we can see actions of type $d$
denoting disturbances, of type $mu$ denoting misuse, and of type $nm$.
Here, \texttt{collision} is an $nm$-typed CF rather than an
unacceptable mishap because we assume to have an airbag \AtomSit{Ab} as a
mitigation.  In \Cref{fig:os-drive}, the action $e^C_{nm}$ leads to a
state with $C$ activated.
\end{example}

\subsection{Which Constraints Reduce the State Space Reachable by
  Endangerments?}
\label{sec:statespacereduction-endg}

The reader might have noticed that the \emph{unconstrained
  composition} of the phase models for $C$, $nC$, $O$, and $W$ in
\Cref{fig:os-drive} would have resulted in a much larger risk
structure.  In fact, we applied several constraints to our model to
avoid the investigation of transitions irrelevant or not meaningful
for designing mechanisms for RAM (\cf \Cref{sec:motivation} and
\Cref{fig:controlledprocess-blackbox}).

For any risk state $\sigma\in\Sigma$ and causal factors
$\cfct_1,\cfct_2$, we can apply the \emph{constraints}:
\begin{itemize}
\item $\cfct_1$ \textbf{requires} $\cfct_2 :\Leftrightarrow$ $\cfct_1$
  can only be activated if $\cfct_2$ has already been activated,
\item $\cfct_1$ \textbf{causes} $\cfct_2 :\Leftrightarrow$ the
  activation of $\cfct_1$ is propagated to $\cfct_2$ (if not yet
  occurred),
\item $\cfct_1$ \textbf{denies} $\cfct_2 :\Leftrightarrow$ the
  activation of $\cfct_1$ denies the activation of $\cfct_2$ (if not
  yet occurred), and
\item $\cfct_1$ \textbf{excludes} $\cfct_2 :\Leftrightarrow$ factor
  $\cfct_1$ renders $\cfct_2$ irrelevant ($\cfct_2$ is deactivated by
  $\cfct_1$).
\end{itemize}
The usage of constraints will generally reduce $\Sigma$.\footnote{This
  goes along the lines of state space reduction in model checking, see
  \egs \cite{Baier2008}.} 
The application of these constraints in a specific causal factor model
is based on, \egs expert knowledge of our safety engineer and on
results of system identification experiments performed by control
engineers.  In particular, \textbf{causes} relationships might have to
be confirmed by, \egs FMEA, \textbf{requires} relationships by, \egs
FTA.

\section{Which Countermeasures Can We Employ At Run-Time?}
\label{sec:runtimemitigation}

In this section, we discuss the risk mitigation part of RAM based on
the hazard analysis part of RAM shown in
\Cref{sec:riskfactorization}.

\subsection{Which Events Decrease Risk? What is a Mitigation Comprised
  of?}
\label{sec:decreasingrisk}

Based on the model in \Cref{sec:terminology}, risk is supposed to be
decreased by \emph{mitigations}.  Mitigations are those actions
performed in $\mathcal{L}$ that result in a mitigation and, finally,
in a deactivation of causal factors, \ies exactly those
risk-decreasing events we want our \emph{safety controller}
(\Cref{fig:controlledprocess-blackbox}) to actuate.

\paragraph{Classifying Mitigations.}
\label{sec:mitigationtaxonomy}

We classify mitigations by the type of action that deactivates a CF.
The following criteria contribute to a taxonomy of mitigations:
\begin{inparaenum}[(i)]
\item the type of endangerment (see \Cref{sec:hazardtaxonomy}),
\item $\ControlLoop$'s mode of operation after mitigation (\egs vehicle
  automation level, speed level), 
\item completion of deactivation of a causal factor at
  run-time. 
\end{inparaenum}
According to this, we distinguish between
\begin{itemize}
\item \emph{fail-safe}, encompassing reactions to AV \emph{failures} by
  \emph{fail-silent} or \emph{fail-operational} behavior,
\item \emph{deescalation}, comprising mechanisms for rejecting
  \emph{disturbances} (\egs stabilization),
\item \emph{protection}, including \emph{failure-independent} mechanisms for
  risk prevention or alleviation of harm,
\item \emph{uncontrolled}, representing mitigation mechanisms
  \emph{not} in control of the AV,
\item \emph{repair and maintenance}, representing mechanisms capable
  of complete deactivation of a CF or restoration of its undesired
  consequences, not necessarily at run-time.
\end{itemize}

\begin{example}
For $C$, our model contains the action $prt^C_{alv}$ which represents
a protection mechanism resulting in the \emph{alleviation} of any
occurrence of $C$ and its consequences.  Finally,
\Cref{fig:os-drive-collision} shows the parts of $\mathcal{L}$
embodying these actions, \egs $\frac{prt^C_{alv}}{\mathit{Ab}}$
indicates that $prt^C_{alv}$ is going to be conducted by an airbag
\AtomSit{Ab} as the alleviating protection mechanism.
\end{example}

\subsection{Which Constraints Shape the State Space Reachable by
  Mitigations?}
\label{sec:statespacereduction-mit}

We can apply \emph{shaping directives} for further
reduction of $\Sigma$:
\begin{itemize}
\item $\cfct$ \textbf{direct}: $\cfct$ can be completely mitigated,
  intentionally at run-time (this corresponds to the action
  $m^{\cfct}_c$ shown in \Cref{fig:hazardphasemodel-generic}),
\item $\cfct$ \textbf{offRepair}: The change of $\cfct$ to the phase
  \emph{inactive} ($0^{\cfct}$) by an action $m_e^{\cfct}$ requires
  putting the AV out of order.
\end{itemize}

\begin{example}
We use the following constraints and shaping directives for $nC$ and
$C$:
\begin{itemize}
\item $nC$ \textbf{requires} $O$: We only consider worlds in which
  \texttt{nearCollision} events require
  \texttt{obstacle\-In\-Trajectory} events to occur in advance of and
  last until at least $nC$.  Note that this constraint inhibits our CF
  model to be used for \emph{passive} \texttt{nearCollision}s.
  This
  issue could be addressed by introducing an extra causal factor.
\item $nC$ \textbf{direct}: Any implementation of a mitigation
  mechanism for $nC$ performs a complete deactivation of $nC$.
\item $C$ \textbf{requires} $nC$: We only consider worlds in
  which any \texttt{nearCollision} event precedes any possible
  \texttt{collision} event.
\item $C$ \textbf{excludes} $nC$: Once a \texttt{collision} event
  happens we stop taking care about any further
  \texttt{near\-Collisions}.
\item $C$ \textbf{offRepair}: Our understanding of \texttt{collisions}
  in road traffic implies that a complete restoration, if possible,
  has to take place off-line, \ies not in the situations shown in
  \Cref{fig:osdecisionprocess-overall}.  This way, the safety
  controller $M$ 
  knows that repair actions have to take place as soon as possible.
  So, ``emergency stop'' and ``limp home'' might become the next
  run-time actions 
  to be performed by $M$.
\end{itemize}
\end{example}

\subsection{Can Endangerments and Mitigations Change the Driving Situation?}
\label{sec:situationchanges}

In general, mitigations change $\Sigma_{\ControlLoop}$, particularly,
the state of the process $P$.  Consequently, \emph{jumps} from
a risk state of one driving situation into a specific risk state of
another situation can occur in many cases.
 
\begin{example}
An \emph{emergency stop} triggered by \texttt{lowOrNo-Fuel} in the
situation \AtomSit{auto}\-\AtomSit{Overtake} would bring the loop into the
situation \AtomSit{halt} or, possibly, into some yet unknown situation
in the region \AtomSit{driveAtLow}\-\AtomSit{Speed}, while carrying on
with further mitigations as required.
\end{example}

\subsection{Towards Assessment by Symbolic Execution}
\label{sec:symbsim}

By \emph{symbolic execution} with \Yap, we can assess our model for
plausibility and refinement.

\begin{table}
  \centering\small
  \begin{tabular}{c>{\itshape}lclcc}
\toprule
Step & \textbf{Driving Situation} & \#CFs & Initial State & \#States & \#Trans.
\\\midrule
1 & start & 

\\
2 & leaveParkingLot & 
3 & $B$ & 9 & 17
\\
3 & driveAtL1Generic & 
10 & $nCoLI$ & 502 & 2165
\\
4 & driveAtL4Generic & 
10 & $FEBWC$ & 1323 & 5726
\\
5 & steerThroughTrafficJam & 
3 & $E$ & 9 & 15
\\
\bottomrule
\end{tabular}

  \caption{Scenario $\rho\in\sem{P0}$ starting from situation
    \AtomSit{start} (\cf$P0$, \Cref{tab:drivingprocess-prodrules}).}
  \label{tab:simrun-start}
\end{table}

\begin{example}
\Cref{tab:simrun-start} shows a driving scenario starting from the
situation \AtomSit{start} with brief statistics per situation
describing the number of causal factors regarded in the current
situation (\#CFs), the initial state to which the \emph{random
  execution} jumps at every step (only listing activated CFs), number
of reachable risk states (particularly RUEs), and number of
transitions (indicating some choice among mitigation options for $M$).
\Cref{tab:simrun-steerThroughTrafficJam} shows another driving
scenario from the situation \AtomSit{steerThroughTrafficJam}.  Both
scenarios were cut after five steps.
\end{example}

\begin{table}
  \centering\small
  \begin{tabular}{c>{\itshape}lclcc}
\toprule
Step & \textbf{Driving Situation} & \#CFs & Initial State & \#States & \#Trans.
\\\midrule
1 & steerThroughTrafficJam & 
3 & $B$ & 14 & 23
\\
2 & driveAtL4Generic & 
10 & $EBOWnAP$ & 2190 & 8432
\\
3 & exitTunnel & 
10 & $BOWrA$ & 3606 & 15337
\\
4 & driveAtL1Generic & 
10 & $EBOWnCI$ & 276 & 868
\\
5 & manuallyOvertake & 
10 & $EO$ & 756 & 2953
\\
\bottomrule
\end{tabular}

  \caption{Scenario $\rho\in\sem{D11}$ starting from situation
    \AtomSit{steerThroughTrafficJam} (\cf $D11$, \Cref{tab:drivingprocess-prodrules}).}
  \label{tab:simrun-steerThroughTrafficJam}
\end{table}

\section{Synergies and Improvements Over Related Work}
\label{sec:impr-over-relat}
\label{sec:relwork}

The questions S2 (Task S), E1-5 (Task E), and M1 (Task M) posed in
\Cref{sec:motivation} are typically addressed by \emph{risk assessment
  techniques} such as, \egs FTA, FMEA \cite{Ericson2015}.  However,
because of the specifics of RAM in AD, the performance of these
techniques heavily depends on loop and causal factor models
available and employed for RAM.  The mentioned techniques rarely
supply specialized and formally verifiable models.
The causal factor model described in \Cref{sec:terminology}
incorporates the principle of ``modeling what can go wrong'' as
employed in practical risk models such as
CORAS~\cite{LundSolhaugStoelen2011}.  Similar to, \egs FTA, FMEA, and
HazOp, the CORAS method itself is not specific to RAM for AD and the
CORAS tool set has not yet been equipped with automation support for
generating and analyzing risk state spaces.  However, we consider a
mapping from CORAS risk models into causal factor models and vice
versa as a valuable extension.
Similarly, the constraints described in
\Cref{sec:statespacereduction-endg,sec:statespacereduction-mit}
resemble some \emph{gates} known from FTA, \egs the AND gate can be
modeled by a \textbf{requires} constraint.

Regarding the \emph{modeling of driving processes}, \cite{TSC2017}
discusses a set of driving situations together with a specific safety
argument based on hazards relevant in each situation.  The approach at
hand allows to transform such a situation taxonomy together with the
arguments into a concise model.

For the \emph{generation of risk state spaces}, we can mention several
approaches:
Volk et al.~\cite{DBLP:conf/safecomp/0001JK16} provide an efficient
technique to generate sparse \textsc{Markov} automata from dynamic
fault trees and for the synthesis of parameters to calculate, \egs the
mean-time-to-failure measure.  Regarding sparseness, \Yap\ does not
yet implement partial-order reduction to a comparable extent
\cite{Baier2008}.
B\"ackstr\"om et al.~\cite{DBLP:conf/safecomp/BackstromBHKK16}
describe an analytic approach to probabilistic, time-limited
reachability of failure states based on continuous-time
\textsc{Markov} chains derived from static and dynamic fault trees.
Similarly, based on stochastic timed automata (STA), 
Kumar and Stoelinga~\cite{DBLP:conf/hase/KumarS17} propose a time- and
budget-limited analysis of a combined model of faults and attacks.
They transform this model using event-class-specific STA templates
rather than having a single generic phase
model~(\cf\Cref{fig:hazardphasemodel-generic}).
However, modeling the controlled process by situations combined with
run-time planning over situation-specific risk state spaces provides
an appropriate context for the employment of these algorithms and for
quantitative assessment.

In the context of \emph{safety standards for unmanned aircrafts},
\cite{Cook2017} sketches a architecture where the main controller
is protected by a safety monitor and a recovery
control. 
Regarding \emph{controller synthesis}, Machin et al.~\cite{Machin2016}
provide a systematic and formal procedure for the synthesis of
intervention rules (i) identified through a UML-based HazOp technique
and (ii) implementing a robot safety monitor.  The rules are
automatically verified 
based on (i) notions of warning and catastrophic states similar to our
generic phase model in \Cref{fig:hazardphasemodel-generic} and (ii) a
measure of permissiveness.  Mitsch and Platzer~\cite{Mitsch2016}
present an intentionally similar approach, however, applicable to
hybrid programs, a more expressive class of control systems.

A more specific line of work deals with \emph{hybrid reachability
  analysis}, \egs \cite{Roehm2016}, where reachable sets are
approximated over metric spaces.  Such approaches are well-suited for
performing optimal and safe \emph{reach-avoid} mitigations for
endangerments like, \egs \texttt{nearCollision}s.  Our approach aims
at a model for developing mitigation strategies for multiple causal
factors and performing mitigations including both, hybrid control
actions and actions performing over non-metric
spaces. 

\section{Discussion}
\label{sec:further-discussion}

This section highlights several aspects of the presented approach to
be pursued by further research.

\subsection{Notes on the Formalization of Constraints}

Let each phase label in \Cref{fig:hazardphasemodel-generic} be an atomic
proposition.  Then, in a 
timed extension of linear time temporal logic
\cite{Manna1991,Manna1995,Koymans1990} and given causal factors
$\cfct_1$ and $\cfct_2$,
\begin{itemize}
\item we translate \textbf{requires} into the formula \hfill
  $\Box[\cfct_1 \rightarrow \blacklozenge_{\leq t}\cfct_2]
  \Leftrightarrow \Box[\neg\cfct_1\mathsf{W}_{\leq t}\,\cfct_2]$,
\item \textbf{causes} into \hfill
  $\Box[\cfct_1\rightarrow\Diamond_{\leq
    t}(\cfct_2\mathsf{U}\neg\cfct_1)]$,
\item \textbf{denies} into \hfill
  $\Box[\cfct_1\rightarrow\Diamond_{\leq
    t}(\neg\cfct_2\mathsf{U}\neg\cfct_1)]$, 
\item and \textbf{excludes} into \hfill
  $\Box[\cfct_1\rightarrow\Diamond_{\leq
    t}(0^{\cfct_2}\mathsf{U}\neg\cfct_1)]$.
\end{itemize}
By having one translation of these constraints into temporal formulas,
we gain a way to translate
\begin{inparaenum}[(i)]
\item any causal factor into a temporal formula by logical conjunction
  ($\land$), and, hence,
\item the causal factor model into a corresponding formula.
\end{inparaenum}
The formula resulting from (ii) characterizes the risk state space
$\Sigma$ without speaking of the action classes described in
\Cref{sec:hazardtaxonomy,sec:mitigationtaxonomy}.

Furthermore, the \textbf{requires} constraint is a relaxed variant of
the \textbf{globally precedes} pattern, \egs discussed in
\cite{Dwyer1999}:
$\cfct_1\;\text{\textbf{requires}}\;\cfct_2 \Leftarrow$
\begin{equation}
  \cfct_2\;\text{\textbf{globally precedes}}\;\cfct_1
  \quad\Leftrightarrow\quad
  \Diamond\cfct_1\rightarrow
  (\neg\cfct_1\mathsf{U}(\cfct_2\land\neg\cfct_1))
  \label{eq:requires2precedes}
\end{equation}
\Cref{eq:requires2precedes} does not hold for runs where $\cfct_1$ and
$\cfct_2$ happen simultaneously and $\cfct_2$ is required to happen.

\subsection{Towards a Research Vision}

Methodically, for RAM, it is important to ask why we need a dedicated
causal factor model for AD?  Well, let us refine
\Cref{thm:toplevelsafetygoal} from \Cref{sec:motivation} by:
\begin{claim}
  \label{thm:toplevelsafetygoal-autonomous}
  An optimal AV controller (AVC) always tries to reach and maintain a
  $b$-safe\footnote{$b$ denotes a maximum ``risk budget'' to encode
    the notion of ``acceptably safe,'' see \Cref{sec:terminology}.}
  \textbf{state} 
  \wrts \textbf{RUEs} recognizable and acted upon by the AVC in any
  known and recognizable \textbf{driving situation}. 
\end{claim} 
From \Cref{thm:toplevelsafetygoal-autonomous} and the answers to
question M4, we can derive planning procedures and verification goals:
are there (i) undesirable mitigations derivable from the model, (ii)
undesirable combinations of situations and mitigations?  How can we
minimize exposure to mitigation-endangerment cycles?  Can we prove
from our model (given it is valid) the (global) existence of an
(effective) \emph{mitigation strategy} incorporating mechanisms from
several parts of $\mathcal{L}$?
Technically, we aim at the instantiation of a monitor-actuator pattern
\cite{Preschern2013c} based on the resulting model of $\mathcal{L}$
depicted in \Cref{fig:controlledprocess-blackbox}.

\subsection{About the Tasks S, E, and M}

For Task S, we need to determine the physical events governing the
driving process and its control, as well as the internal events in the
implementation of the controller.
For Task E, a predictive model of operational risk in any driving
situation would be of great help.  However, we only made a first step
into the direction of discrete event model-predictive control.
For Task M, our mitigation strategy always depends on the mechanisms
and abilities feasible and controllable in $\mathcal{L}$.

\subsection{Notes on Terminology, Assumptions, and Limitations}

Our assumption for FAD, worth mentioning, is that \emph{functional
  safety} of the controller is not considered in separation of
\emph{overall safety} of the control loop, which goes along the lines
of \cite{Leveson2012}.
It might seem cumbersome to speak of ``rare'' \emph{undesired
  events}.  However, \emph{frequent} undesired events might stem from
systematic defects or disturbances in the control loop for which
general design-time measures might be desired before applying the
discussed RAM approach.

We excluded the consideration of probabilities of causal factors and
transitions in the driving process.  However, stochastic reasoning
helps in quantitative approximation for any planning algorithm
employed for \Cref{thm:toplevelsafetygoal-autonomous}.
Moreover, probabilities will play a role in quantifying to which
extent the model from \Cref{eq:proccalc} represents the controlled
process.
In \Cref{sec:mitigationtaxonomy,sec:hazardtaxonomy}, we left
classifications of endangerments and mitigations shallow, though,
\Yap\ already incorporates a slightly more elaborate taxonomy.

\subsection{Summary and Future Work}
\label{sec:conclusions}

In this paper, we discussed \emph{first steps} of an approach to risk
analysis and run-time mitigation (RAM) suited for
\begin{inparaenum}[(i)]
\item the investigation of the driving process (Task S, S1-3),
\item capturing multiple causal factors forming a set of
  rare undesired events in the risk state space (Task E, E1-5), and
\item designing run-time mitigations to traverse the risk state space
  in a manner to establish~\Cref{thm:toplevelsafetygoal-autonomous,thm:toplevelsafetygoal}
  (Task M, M1-4).
\end{inparaenum}
Our model supports \emph{safety engineering decision making} and the
transfer of such decisions into a \emph{safety controller}.  This
safety controller is supposed to implement the monitor in
\Cref{fig:controlledprocess-blackbox} as well as a \emph{strategic
  mitigation planner} for run-time risk mitigation.

\paragraph{Next Steps.}
For the formalization of the discussed concepts, one next step is to
provide \textsc{Markov} decision process semantics for the process
model described in \Cref{sec:sitmodel} and to map risk structures
(\ies the situation-specific LTS models described in
\Cref{sec:terminology,sec:riskfactorization}) into equivalent
\textsc{Kripke} structures.
Extending these structures, we aim to equip our RAM approach for
continuous update of real-time determined weights.
Next, \Yap\ which demonstrates part of the automated analyses for RAM
(\ies state space generation) has to be refined.%
Beyond these steps, we aim at the enhancement of the presented
approach towards a controller design method integrating run-time risk
mitigation for automated individual and collective driving.

\paragraph*{Acknowledgments.}
I want to thank several engineers from German car makers and suppliers
with whom I had many insightful discussions on this line of work.  My
sincere gratitude goes to my mentor and former advisor Manfred Broy
for always putting highly valuable critique in a nutshell.  I am
grateful for a supportive work atmosphere in our collaboration with
industry established by my project leader Stefan Kugele who has
accompanied me with exciting previous work on this topic.

\bibliographystyle{eptcs}
\bibliography{references}

\begin{thebibliography}{10}
\providecommand{\bibitemdeclare}[2]{}
\providecommand{\surnamestart}{}
\providecommand{\surnameend}{}
\providecommand{\urlprefix}{Available at }
\providecommand{\url}[1]{\texttt{#1}}
\providecommand{\href}[2]{\texttt{#2}}
\providecommand{\urlalt}[2]{\href{#1}{#2}}
\providecommand{\doi}[1]{doi:\urlalt{http://dx.doi.org/#1}{#1}}
\providecommand{\bibinfo}[2]{#2}

\bibitemdeclare{article}{Dowell1998}
\bibitem{Dowell1998}
\bibinfo{author}{\surnamestart {Arthur M. Dowell III}\surnameend}
  (\bibinfo{year}{1998}): \emph{\bibinfo{title}{Layer of protection analysis
  for determining safety integrity level}}.
\newblock {\sl \bibinfo{journal}{ISA Transactions}}
  \bibinfo{volume}{37}(\bibinfo{number}{3}), pp. \bibinfo{pages}{155 -- 165},
  \doi{10.1016/S0019-0578(98)00018-4}.

\bibitemdeclare{inproceedings}{DBLP:conf/safecomp/BackstromBHKK16}
\bibitem{DBLP:conf/safecomp/BackstromBHKK16}
\bibinfo{author}{Ola \surnamestart B{\"{a}}ckstr{\"{o}}m\surnameend},
  \bibinfo{author}{Yuliya \surnamestart Butkova\surnameend},
  \bibinfo{author}{Holger \surnamestart Hermanns\surnameend},
  \bibinfo{author}{Jan \surnamestart Krc{\'{a}}l\surnameend} \&
  \bibinfo{author}{Pavel \surnamestart Krc{\'{a}}l\surnameend}
  (\bibinfo{year}{2016}): \emph{\bibinfo{title}{Effective Static and Dynamic
  Fault Tree Analysis}}.
\newblock In \bibinfo{editor}{Skavhaug} et~al.  \cite{DBLP:conf/safecomp/2016},
  pp. \bibinfo{pages}{266--280}, \doi{10.1007/978-3-319-45477-1\_21}.

\bibitemdeclare{book}{Baier2008}
\bibitem{Baier2008}
\bibinfo{author}{Christel \surnamestart Baier\surnameend} \&
  \bibinfo{author}{Joost-Pieter \surnamestart Katoen\surnameend}
  (\bibinfo{year}{2008}): \emph{\bibinfo{title}{Principles of Model Checking}}.
\newblock \bibinfo{publisher}{MIT Press}.

\bibitemdeclare{book}{Broy2001}
\bibitem{Broy2001}
\bibinfo{author}{Manfred \surnamestart Broy\surnameend} \&
  \bibinfo{author}{Ketil \surnamestart St{\o}len\surnameend}
  (\bibinfo{year}{2001}): \emph{\bibinfo{title}{Specification and Development
  of Interactive Systems: \textsc{Focus} on Streams, Interfaces, and
  Refinement}}.
\newblock \bibinfo{publisher}{Springer}, \bibinfo{address}{Berlin},
  \doi{10.1007/978-1-4613-0091-5}.

\bibitemdeclare{incollection}{Cook2017}
\bibitem{Cook2017}
\bibinfo{author}{Stephen~P. \surnamestart Cook\surnameend}
  (\bibinfo{year}{2017}): \emph{\bibinfo{title}{An ASTM Standard for Bounding
  Behavior of Adaptive Algorithms for Unmanned Aircraft Operations (Invited)}}.
\newblock \bibinfo{series}{AIAA SciTech Forum}, \bibinfo{publisher}{American
  Institute of Aeronautics and Astronautics}, \doi{10.2514/6.2017-0881}.

\bibitemdeclare{inproceedings}{Dwyer1999}
\bibitem{Dwyer1999}
\bibinfo{author}{Matthew~B. \surnamestart Dwyer\surnameend},
  \bibinfo{author}{G.~S. \surnamestart Avrunin\surnameend} \&
  \bibinfo{author}{J.~C. \surnamestart Corbett\surnameend}
  (\bibinfo{year}{1999}): \emph{\bibinfo{title}{Patterns in property
  specifications for finite-state verification}}.
\newblock In: {\sl \bibinfo{booktitle}{ICSE}}, pp. \bibinfo{pages}{411--20},
  \doi{10.1109/icse.1999.841031}.

\bibitemdeclare{inproceedings}{Eastwood2013}
\bibitem{Eastwood2013}
\bibinfo{author}{R.~\surnamestart Eastwood\surnameend}, \bibinfo{author}{T.P.
  \surnamestart Kelly\surnameend}, \bibinfo{author}{R.D. \surnamestart
  Alexander\surnameend} \& \bibinfo{author}{E.~\surnamestart Landre\surnameend}
  (\bibinfo{year}{2013}): \emph{\bibinfo{title}{Towards a safety case for
  runtime risk and uncertainty management in safety-critical systems}}.
\newblock In: {\sl \bibinfo{booktitle}{System Safety Conference incorporating
  the Cyber Security Conference 2013, 8th IET International}}, pp.
  \bibinfo{pages}{1--6}, \doi{10.1049/cp.2013.1713}.

\bibitemdeclare{book}{Ericson2015}
\bibitem{Ericson2015}
\bibinfo{author}{Clifton~A. \surnamestart Ericson\surnameend}
  (\bibinfo{year}{2015}): \emph{\bibinfo{title}{Hazard Analysis Techniques for
  System Safety}}, \bibinfo{edition}{2nd} edition.
\newblock \bibinfo{publisher}{Wiley}.

\bibitemdeclare{phdthesis}{Gleirscher2014a}
\bibitem{Gleirscher2014a}
\bibinfo{author}{Mario \surnamestart Gleirscher\surnameend}
  (\bibinfo{year}{2014}): \emph{\bibinfo{title}{Behavioral Safety of Technical
  Systems}}.
\newblock \bibinfo{type}{Dissertation}, \bibinfo{school}{Technische
  Universit{\"a}t M{\"u}nchen}, \doi{10.13140/2.1.3122.7688}.

\bibitemdeclare{inproceedings}{Gleirscher2017b}
\bibitem{Gleirscher2017b}
\bibinfo{author}{Mario \surnamestart Gleirscher\surnameend} \&
  \bibinfo{author}{Stefan \surnamestart Kugele\surnameend}
  (\bibinfo{year}{2017}): \emph{\bibinfo{title}{Defining Risk States in
  Autonomous Road Vehicles}}.
\newblock In: {\sl \bibinfo{booktitle}{High Assurance Systems Engineering
  (HASE), 18th Int. Symp.}}, pp. \bibinfo{pages}{112--115},
  \doi{10.1109/hase.2017.14}.

\bibitemdeclare{inproceedings}{Gleirscher2017-NFM}
\bibitem{Gleirscher2017-NFM}
\bibinfo{author}{Mario \surnamestart Gleirscher\surnameend} \&
  \bibinfo{author}{Stefan \surnamestart Kugele\surnameend}
  (\bibinfo{year}{2017}): \emph{\bibinfo{title}{From Hazard Analysis to Hazard
  Mitigation Planning: The Automated Driving Case}}.
\newblock In \bibinfo{editor}{C.~Barrett \surnamestart et~al.\surnameend},
  editor: {\sl \bibinfo{booktitle}{{NASA} Formal Methods ({NFM}) -- 9th Int.
  Symp., Proceedings}}, {\sl \bibinfo{series}{LNCS}} \bibinfo{volume}{10227},
  \bibinfo{publisher}{Springer, Berlin/New York}, pp.
  \bibinfo{pages}{310--326}, \doi{10.1007/978-3-319-57288-8\_23}.

\bibitemdeclare{article}{Guiochet2017}
\bibitem{Guiochet2017}
\bibinfo{author}{Jeremie \surnamestart Guiochet\surnameend},
  \bibinfo{author}{Mathilde \surnamestart Machin\surnameend} \&
  \bibinfo{author}{Helene \surnamestart Waeselynck\surnameend}
  (\bibinfo{year}{2017}): \emph{\bibinfo{title}{Safety-critical Advanced
  Robots: A Survey}}.
\newblock {\sl \bibinfo{journal}{Robots and Autonomous Systems}},
  \doi{10.1016/j.robot.2017.04.004}.

\bibitemdeclare{book}{Hoare1985}
\bibitem{Hoare1985}
\bibinfo{author}{Charles A.~R. \surnamestart Hoare\surnameend}
  (\bibinfo{year}{1985}): \emph{\bibinfo{title}{Communicating Sequential
  Processes}}, \bibinfo{edition}{1st} edition.
\newblock \bibinfo{series}{Int. Series in Comp. Sci.},
  \bibinfo{publisher}{Prentice-Hall}.

\bibitemdeclare{inproceedings}{Koopman2016}
\bibitem{Koopman2016}
\bibinfo{author}{Phil \surnamestart Koopman\surnameend} \&
  \bibinfo{author}{Michael \surnamestart Wagner\surnameend}
  (\bibinfo{year}{2016}): \emph{\bibinfo{title}{Challenges in Autonomous
  Vehicle Testing and Validation}}.
\newblock In: {\sl \bibinfo{booktitle}{SAE World Congress}},
  \doi{10.4271/2016-01-0128}.

\bibitemdeclare{article}{Koymans1990}
\bibitem{Koymans1990}
\bibinfo{author}{Ron \surnamestart Koymans\surnameend} (\bibinfo{year}{1990}):
  \emph{\bibinfo{title}{Specifying real-time properties with metric temporal
  logic}}.
\newblock {\sl \bibinfo{journal}{Real-Time Syst.}}
  \bibinfo{volume}{2}(\bibinfo{number}{4}), pp. \bibinfo{pages}{255--99},
  \doi{10.1007/bf01995674}.

\bibitemdeclare{inproceedings}{DBLP:conf/hase/KumarS17}
\bibitem{DBLP:conf/hase/KumarS17}
\bibinfo{author}{Rajesh \surnamestart Kumar\surnameend} \&
  \bibinfo{author}{Mari{\"{e}}lle \surnamestart Stoelinga\surnameend}
  (\bibinfo{year}{2017}): \emph{\bibinfo{title}{Quantitative Security and
  Safety Analysis with Attack-Fault Trees}}.
\newblock In: {\sl \bibinfo{booktitle}{18th {IEEE} International Symposium on
  High Assurance Systems Engineering, {HASE} 2017, Singapore, January 12-14,
  2017}}, \bibinfo{publisher}{{IEEE}}, pp. \bibinfo{pages}{25--32},
  \doi{10.1109/HASE.2017.12}.

\bibitemdeclare{article}{DBLP:journals/tse/Lamport77}
\bibitem{DBLP:journals/tse/Lamport77}
\bibinfo{author}{Leslie \surnamestart Lamport\surnameend}
  (\bibinfo{year}{1977}): \emph{\bibinfo{title}{Proving the Correctness of
  Multiprocess Programs}}.
\newblock {\sl \bibinfo{journal}{{IEEE} Trans. Software Eng.}}
  \bibinfo{volume}{3}(\bibinfo{number}{2}), pp. \bibinfo{pages}{125--43},
  \doi{10.1109/TSE.1977.229904}.

\bibitemdeclare{book}{Leveson2012}
\bibitem{Leveson2012}
\bibinfo{author}{Nancy~Gail \surnamestart Leveson\surnameend}
  (\bibinfo{year}{2012}): \emph{\bibinfo{title}{Engineering a Safer World:
  Systems Thinking Applied to Safety}}.
\newblock \bibinfo{series}{Engineering Systems}, \bibinfo{publisher}{MIT
  Press}.

\bibitemdeclare{book}{LundSolhaugStoelen2011}
\bibitem{LundSolhaugStoelen2011}
\bibinfo{author}{Mass~Soldal \surnamestart Lund\surnameend},
  \bibinfo{author}{Bj{\o}rnar \surnamestart Solhaug\surnameend} \&
  \bibinfo{author}{Ketil \surnamestart St{\o}len\surnameend}
  (\bibinfo{year}{2011}): \emph{\bibinfo{title}{Model-Driven Risk Analysis: The
  {CORAS} Approach}}, \bibinfo{edition}{1st} edition.
\newblock \bibinfo{publisher}{Springer}, \doi{10.1007/978-3-642-12323-8}.

\bibitemdeclare{book}{Lunze2010}
\bibitem{Lunze2010}
\bibinfo{author}{Jan \surnamestart Lunze\surnameend} (\bibinfo{year}{2010}):
  \emph{\bibinfo{title}{Regelungstechnik 1: Systemtheoretische Grundlagen,
  Analyse und Entwurf einschleifiger Regelungen}}, \bibinfo{edition}{8th}
  edition.
\newblock \bibinfo{series}{Lehrbuch}, \bibinfo{publisher}{Springer},
  \doi{10.1007/978-3-642-13808-9}.

\bibitemdeclare{article}{Machin2016}
\bibitem{Machin2016}
\bibinfo{author}{Mathilde \surnamestart Machin\surnameend},
  \bibinfo{author}{J\'{e}r\'{e}mie \surnamestart Guiochet\surnameend},
  \bibinfo{author}{H\'{e}l\`{e}ne \surnamestart Waeselynck\surnameend},
  \bibinfo{author}{Jean-Paul \surnamestart Blanquart\surnameend},
  \bibinfo{author}{Matthieu \surnamestart Roy\surnameend} \&
  \bibinfo{author}{Lola \surnamestart Masson\surnameend}
  (\bibinfo{year}{2016}): \emph{\bibinfo{title}{{SMOF} -- A {S}afety
  {MO}nitoring {F}ramework for Autonomous Systems}} \bibinfo{volume}{99}, pp.
  \bibinfo{pages}{1--14}.
\newblock \doi{10.1109/tsmc.2016.2633291}.

\bibitemdeclare{book}{Manna1991}
\bibitem{Manna1991}
\bibinfo{author}{Zohar \surnamestart Manna\surnameend} \& \bibinfo{author}{Amir
  \surnamestart Pnueli\surnameend} (\bibinfo{year}{1991}):
  \emph{\bibinfo{title}{The Temporal Logic of Reactive and Concurrent Systems:
  Specification}}, \bibinfo{edition}{1st} edition.
\newblock \bibinfo{publisher}{Springer}.

\bibitemdeclare{book}{Manna1995}
\bibitem{Manna1995}
\bibinfo{author}{Zohar \surnamestart Manna\surnameend} \& \bibinfo{author}{Amir
  \surnamestart Pnueli\surnameend} (\bibinfo{year}{1995}):
  \emph{\bibinfo{title}{Temporal Verification of Reactive Systems: Safety}},
  \bibinfo{edition}{1st} edition.
\newblock \bibinfo{publisher}{Springer}, \doi{10.1007/978-1-4612-4222-2}.

\bibitemdeclare{book}{Milner1995}
\bibitem{Milner1995}
\bibinfo{author}{Robin \surnamestart Milner\surnameend} (\bibinfo{year}{1995}):
  \emph{\bibinfo{title}{Communication and Concurrency}}.
\newblock \bibinfo{series}{International Series in Computer Science},
  \bibinfo{publisher}{Prentice Hall}.

\bibitemdeclare{article}{Mitsch2016}
\bibitem{Mitsch2016}
\bibinfo{author}{Stefan \surnamestart Mitsch\surnameend} \&
  \bibinfo{author}{Andr\'{e} \surnamestart Platzer\surnameend}
  (\bibinfo{year}{2016}): \emph{\bibinfo{title}{ModelPlex: Verified Runtime
  Validation of Verified Cyber-Physical System Models}}.
\newblock \doi{10.1007/978-3-319-11164-3\_17}.

\bibitemdeclare{techreport}{Nielsen1971}
\bibitem{Nielsen1971}
\bibinfo{author}{D.S. \surnamestart Nielsen\surnameend} (\bibinfo{year}{1971}):
  \emph{\bibinfo{title}{The cause/consequence diagram method as basis for
  quantitative accident analysis}}.
\newblock \bibinfo{type}{Technical Report} \bibinfo{number}{RISO-M-1374},
  \bibinfo{institution}{Danish Atomic Energy Commission}.

\bibitemdeclare{techreport}{ORADC2016}
\bibitem{ORADC2016}
\bibinfo{author}{\surnamestart {On-Road Automated Driving
  Committee}\surnameend} (\bibinfo{year}{2016}): \emph{\bibinfo{title}{Taxonomy
  and Definitions for Terms Related to Driving Automation Systems for On-Road
  Motor Vehicles}}.
\newblock \bibinfo{type}{Technical Report} \bibinfo{number}{SAE J 3016},
  \bibinfo{institution}{SAE International}, \doi{10.4271/j3016_201609}.

\bibitemdeclare{inproceedings}{Preschern2013c}
\bibitem{Preschern2013c}
\bibinfo{author}{Christopher \surnamestart Preschern\surnameend},
  \bibinfo{author}{Nermin \surnamestart Kajtazovic\surnameend} \&
  \bibinfo{author}{Christian \surnamestart Kreiner\surnameend}
  (\bibinfo{year}{2013}): \emph{\bibinfo{title}{Building a safety architecture
  pattern system}}.
\newblock In \bibinfo{editor}{Uwe \surnamestart van Heesch\surnameend} \&
  \bibinfo{editor}{Christian \surnamestart Kohls\surnameend}, editors: {\sl
  \bibinfo{booktitle}{Proceedings of the 18th European Conference on Pattern
  Languages of Programs (EuroPLoP), Irsee, Germany, July 10-14, 2013}},
  \bibinfo{publisher}{{ACM}}, p.~\bibinfo{pages}{17},
  \doi{10.1145/2739011.2739028}.

\bibitemdeclare{inbook}{Roehm2016}
\bibitem{Roehm2016}
\bibinfo{author}{Hendrik \surnamestart Roehm\surnameend}, \bibinfo{author}{Jens
  \surnamestart Oehlerking\surnameend}, \bibinfo{author}{Thomas \surnamestart
  Heinz\surnameend} \& \bibinfo{author}{Matthias \surnamestart
  Althoff\surnameend} (\bibinfo{year}{2016}): \emph{\bibinfo{title}{STL Model
  Checking of Continuous and Hybrid Systems}}, pp. \bibinfo{pages}{412--27}.
\newblock \bibinfo{publisher}{Springer}, \doi{10.1007/978-3-319-46520-3\_26}.

\bibitemdeclare{proceedings}{DBLP:conf/safecomp/2016}
\bibitem{DBLP:conf/safecomp/2016}
\bibinfo{editor}{Amund \surnamestart Skavhaug\surnameend},
  \bibinfo{editor}{J{\'{e}}r{\'{e}}mie \surnamestart Guiochet\surnameend} \&
  \bibinfo{editor}{Friedemann \surnamestart Bitsch\surnameend}, editors
  (\bibinfo{year}{2016}): \emph{\bibinfo{title}{Computer Safety, Reliability,
  and Security - 35th International Conference, {SAFECOMP} 2016, Trondheim,
  Norway, September 21-23, 2016, Proceedings}}. {\sl \bibinfo{series}{Lecture
  Notes in Computer Science}} \bibinfo{volume}{9922},
  \bibinfo{publisher}{Springer}, \doi{10.1007/978-3-319-45477-1}.

\bibitemdeclare{article}{Svedung2002}
\bibitem{Svedung2002}
\bibinfo{author}{I.~\surnamestart Svedung\surnameend} \&
  \bibinfo{author}{J.~\surnamestart Rasmussen\surnameend}
  (\bibinfo{year}{2002}): \emph{\bibinfo{title}{Graphic representation of
  accident scenarios: Mapping system structure and the causation of
  accidents}}.
\newblock {\sl \bibinfo{journal}{Safety Science}}
  \bibinfo{volume}{40}(\bibinfo{number}{5}), pp. \bibinfo{pages}{397--417},
  \doi{10.1016/s0925-7535(00)00036-9}.

\bibitemdeclare{techreport}{TSC2017}
\bibitem{TSC2017}
\bibinfo{author}{\surnamestart {Transport Systems Catapult}\surnameend}
  (\bibinfo{year}{2017}): \emph{\bibinfo{title}{Taxonomy of Scenarios for
  Automated Driving}}.
\newblock \bibinfo{type}{Technical Report}, \bibinfo{institution}{Transport
  Systems Catapult}.

\bibitemdeclare{inproceedings}{DBLP:conf/fortest/Tretmans08}
\bibitem{DBLP:conf/fortest/Tretmans08}
\bibinfo{author}{Jan \surnamestart Tretmans\surnameend} (\bibinfo{year}{2008}):
  \emph{\bibinfo{title}{Model Based Testing with Labelled Transition Systems}}.
\newblock In: {\sl \bibinfo{booktitle}{Formal Methods and Testing}}, pp.
  \bibinfo{pages}{1--38}, \doi{10.1007/978-3-540-78917-8\_1}.

\bibitemdeclare{techreport}{USDOT2016}
\bibitem{USDOT2016}
\bibinfo{author}{\surnamestart {U.S. Department of Transportation}\surnameend}
  (\bibinfo{year}{2016}): \emph{\bibinfo{title}{Federal Automated Vehicles
  Policy}}.
\newblock \bibinfo{type}{Technical Report}, \bibinfo{institution}{U.S.
  Department of Transportation}.

\bibitemdeclare{inproceedings}{DBLP:conf/safecomp/0001JK16}
\bibitem{DBLP:conf/safecomp/0001JK16}
\bibinfo{author}{Matthias \surnamestart Volk\surnameend},
  \bibinfo{author}{Sebastian \surnamestart Junges\surnameend} \&
  \bibinfo{author}{Joost{-}Pieter \surnamestart Katoen\surnameend}
  (\bibinfo{year}{2016}): \emph{\bibinfo{title}{Advancing Dynamic Fault Tree
  Analysis - Get Succinct State Spaces Fast and Synthesise Failure Rates}}.
\newblock In \bibinfo{editor}{Skavhaug} et~al.  \cite{DBLP:conf/safecomp/2016},
  pp. \bibinfo{pages}{253--265}, \doi{10.1007/978-3-319-45477-1\_20}.

\end{thebibliography}
\end{document}